\documentstyle[preprint,aps]{revtex}
\begin{document}

\preprint{BGU PH-95/09\\
hep-ph/9510444}
\draft

\title{Light$\leftrightarrow$Heavy Symmetry:\\
Geometric Mass Hierarchy for Three Families}

\author{Aharon Davidson\thanks{davidson@bgumail.bgu.ac.il} and
Tomer Schwartz\thanks{shwartz@bgumail.bgu.ac.il}}
\address{ Physics Department,
Ben-Gurion University of the Negev,
Beer-Sheva 84105, Israel}

\author{Kameshwar C. Wali\thanks{wali@suhep.phy.syr.edu}}
\address{Physics Deparment, Syracuse University,
Syracuse NY, 13244-1130}
\maketitle

\begin{abstract}
The Universal Seesaw pattern coupled with a
Light$\leftrightarrow$Heavy symmetry principle
leads to the Diophantine equation $\displaystyle N = \sum_{i=1}^Nn_i$,
where $n_i\geq 0$ and distinct. Its unique non-trivial solution
$(3=0+1+2)$ gives rise to the geometric mass hierarchy  $m_W$,
$m_W\epsilon$, $m_W\epsilon^2$  for $N=3$ fermion families.
This is realized in a model where the hybrid (yet
Up$\leftrightarrow$Down symmetric) quark mass
relations $m_d m_t \approx m_c^2\leftrightarrow m_u m_b \approx m_s^2 $
play a crucial role in expressing the CKM
mixings in terms of simple mass ratios, notably
$\sin\theta_C \approx {m_c\over m_b}$.
\end{abstract}


\pagebreak

The physics which governs the Yukawa sector is rooted beyond the standard
$SU(3)\ast SU(2) \ast U(1) $ electro/weak theory. Despite the fact that quark
(lepton) masses and mixings are fairly well known, their observed mass
hierarchy
does not have at the moment a solid theoretical ground, not even a reasonable
empirical formulation. The finest ideas around have so far fallen short of
decoding the three family Fermi Puzzle. Supersymmetry (R-symmetry included)
does not allow the same scalar couple in both Up and Down sectors, but
otherwise leaves the Yukawa couplings fully arbitrary. Grand unification (GUT),
attainable at the single-family level, has consequently little to
say about the horizontal features of the fermion spectrum.
And superstring theory, loaded with self-consistency and inspiration,
has yet to become the theory of everything (TOE). In
fact, it appears as though one must patiently wait the
entry of (quantum) gravity into the game. In the meantime, it always
makes sense to pay serious attention to the accumulating experimental clues.

In this letter, we introduce
the so-called Light-Heavy symmetry in an attempt to account for the
observed
geometric Fermi mass hierarchy. The natural
framework to host such a
symmetry principle is the Universal Seesaw (US) model \onlinecite{US}. In this
unifiable model,
a Froggatt-Nielsen-type \onlinecite{Froggatt-Nielsen} mechanism was implemented
in a universal manner, without
appealing to any hierarchy among the non-vanishing Yukawa couplings. It was
originally designed, using a simplified 'square root' Higgs system, to actually
predict the Gell-Mann-Yanagida \onlinecite{Gell-Mann-Yanagida}
$m_\nu \ll m_e$ once having accounted for $m_{e,u,d} \ll m_W$. In what follows,
we shall show that the pattern of the US mechanism combined with the
Light-Heavy
symmetry idea points uniquely to $N=3$ fermion families, and dictates a
geometric hierarchy among their masses. This is achieved without upsetting the
symmetric interplay between the Up and the Down sectors. We present the
arguments in three steps:

\noindent \underline{Step I:} Consider a typical US mass sub-matrix of the form
\begin{equation}
\left(
\begin{array}{ccccccc}
 &\, &\, &\, &\, &\,m&\\
 &\, &\, &\, &\,\chi&\,M&\\
 &\, &\, &\,.&\,M&\, &\\
 &\, &\,.&\,.&\, &\, &\\
 &\,\chi&\,.&\, &\, &\, &\\
\chi&\,M&\, &\, &\, &\, &\\
\end{array}
\right)
{}~,
\label{typical}
\end{equation}
where $m$ denotes the electro/weak mass scale. $\chi$ and $M$ are the two
$SU(3) \ast SU(2) \ast U(1)$-invariant heavy mass scales,
$m \ll \chi \ll M$, whose ratio defines the US hierarchy parameter
$\epsilon \equiv {\chi \over M}$. If $n$ seesaw partners (weak singlets
with matching $SU(3)_c \ast U(1)_{e.m}$ assignments) are involved,
the lightest eigenvalue of the above
$(n+1)$-dimensional sub-matrix is of order $m\epsilon^n$.

\noindent \underline{Step II:} Requiring an arbitrary
mass hierarchy among N standard families, to be precise,
$m\epsilon^{n_1}, m\epsilon^{n_2}, ..., m\epsilon^{n_N}$
such that $n_i\neq{n_j}$ if $i\neq{j}$, simply
means introducing a total number
$n_1+n_2+...+n_N$ of exotic seesaw families into the theory.
Further, if we impose a Light-Heavy symmetry principle to pair
one seesaw partner F with each standard fermion f, we obtain
the Diophantine equation
\begin{equation}
n_1+n_2+...+n_N=N
{}~,
\end{equation}
to be satisfied by a set $n_i (i=1,...,N)$ of distinct non-negative integers.

\noindent \underline{Step III:} The one-family solution
($N=1; n_1=1$) constitutes the
original US model. This solution as well as the two-family solution
($N=2; n_1=0, n_2=2$) are nothing but the building blocks of the
\underline{only} non-trivial solution ($N=3; n_1=0, n_2=1, n_3=2$). Thus,
not only have we correlated the total number $N=3$ of families with the
Fermi mass hierarchy, but we
can also infer that

\noindent (i) Owing to $n_1=0$, the heaviest standard
family necessarily picks up the electro/weak mass scale
(an encouraging result given the top mass
$m_t\approx{2m_W}$), and

\noindent (ii) Owing to $n_1+n_3=2n_2$, the hierarchy is necessarily geometric.

We now proceed to construct explicitly the three-family model. Our
6x6 mass texture consists of the three blocks associated with
$n_1=0, n_2=1, n_3=2$ of the prototype form (\ref{typical}). But how are the
remainder entries (to be referred to as block mixings) to be decided?
After all, they can have
undesirable (as well as desirable) consequences in the low-energy
regime. A first reasonable criterion would be that the new entries
should not introduce leading order corrections to the hierarchical
eigenvalues  $m_W, m_W\epsilon, m_W\epsilon^2$. A second consideration
would be whether the resulting matrix has any symmetries left and
whether it produces the right order of mixings. To illustrate these
points, let us consider the two-family 3x3 sub-texture
\begin{equation}
\left(
\begin{array}{cc|c}
0& \tilde{m}& m\\
\hline
0& m& 0\\
\chi& M& \tilde{\chi}\\
\end{array}
\right)
{}~,
\label{mixing1}
\end{equation}
where the carefully located zeroes assure that in the limit
$M \rightarrow \infty$ there is no vestige of the second family.
Next, if both block mixings
$\tilde{m}$ and $\tilde{\chi}$ are non-vanishing, there is no possibility
for a remnant symmetry that predicts this pattern. Consequently, at
least one of them should be zero.
If $\tilde{\chi}=0$, we can show that the
mixing angle $\theta_L$ between the two light
left-handed fermions is
$\theta_L\sim{\cal O}(\epsilon^2)$.
If, on the other hand, $\tilde{m}=0$, we derive
$\theta_L\sim{\cal O}(\epsilon)$.
The latter choice is preferred if one wants to avoid the situation
where the nearest neighbor mixings get doubly-suppressed.
One need not be discouraged though by the fact that, unlike in
the conventional Weinberg-Fritzsch
prescription \onlinecite{Weinberg-Fritzsch} for relatively large
(square-mass-ratio) mixings
$\sim \sqrt{\lambda_i \over \lambda_j}$ (given $\lambda_i \ll \lambda_j$),
our approach can only offer apparently small (mass-ratio) mixings
$\sim {\lambda_i \over \lambda_j}$.

Using similar arguments, we are led to a unique 6x6 mixing pattern
for three standard families, and would like
to support it by a specific Light-Heavy symmetry realization.
We aim towards the Up-Down symmetric spectrum
\begin{equation}
\begin{array}{lclc}
m_t\approx m & ,& m_b\approx n& ,\\
m_c\approx m\mid {x \over M}\mid & ,& m_s\approx n\mid {y \over M}\mid& ,\\
m_u\approx n\mid {y \over M}\mid^2& ,&m_d\approx m\mid {x \over M}\mid^2& ,\\
\end{array}
\label{spectrum}
\end{equation}
supposed to hold at some common (yet to be specified)
mass scale. The reasons for considering such a \underline{hybrid}
geometric mass hierarchy are threefold:
(i) The above alternative seems to be numerically preferred,
(ii) Since $m_em_\tau\approx m^2_\mu$ relation is badly violated in
the charged lepton sector, there is no reason for insisting on
$m_um_t\approx m^2_c\leftrightarrow m_dm_b\approx m^2_s$
in the quark sector, and
(iii) $m \over n$  may serve to enhance the Cabibbo angle (that is
$\theta_c\sim{m\over n}{x\over M}$, rather than
$\theta_c\sim{x\over M},{y\over M}$).
\noindent Assuming that all non-vanishing Yukawa
couplings are of the same order of magnitude, $m\equiv <\phi_1>$ and
$n\equiv <\phi_2>$ are essentially the VEVs of the two Higgs doublets involved,
whereas $x\equiv < \chi_1>$ and $y\equiv< \chi_2>$  are the VEVs of
their singlet companions, respectively. Recall
that doublets and singlet Higgs scalars play a perfectly
symmetric Yukawa role \onlinecite{US} in the US mechanism.

To derive the above pattern, we appeal to a horizontal $U(1)_Q$
global symmetry (soon to be discretized on electro/weak grounds).
A central role in our analysis is played by the 6x6 matrix
\begin{equation}
Q^{(\Psi)}_{ij}\equiv Q(\Psi_{iL})-Q(\Psi_{iR}) ~,\label{defQ}
\end{equation}
defined for each electrically-charged fermion sector.
Denoting the Q-charges of the
scalars by $Q(\phi_1)\equiv \alpha$, $Q(\phi_2)\equiv \beta$,
$Q(\chi_1)\equiv a$, $Q(\chi_2)\equiv b$, the rules of the game are quite
simple:
\begin{equation}
\begin{array}{cccc}
\text{ for }&\Psi_L=u_L :&Q^{up}_{ij}=\alpha,\beta \Rightarrow M^{up}_{ij}
= m,n&, \\
\text{ for }&\Psi_L=d_L :&Q^{down}_{ij}=-\beta,-\alpha \Rightarrow
M^{down}_{ij}
=n^*,m^*&, \\
\text{ for }&\Psi_L=F_L :&Q_{ij}=a,b,-a,-b \Rightarrow M_{ij}
=x,y,x^*,y^*&. \\
\end{array}
\end{equation}
Notice the fine differences of the $SU(3)\ast SU(2) \ast U(1)$
restrictions on the Higgs singlet versus doublet couplings.
For example, $\overline{U_L}\chi u_R$ and $\overline{U_L}\chi^{\dag} u_R$
are both allowed. However, in the case of ordinary quarks,
$\overline{q_L}\phi u_R$ and
$\overline{q_L}\phi^{\dag} d_R$ are allowed,
$\overline{q_L}\phi^{\dag} u_R$ and $\overline{q_L}\phi d_R$ are forbidden.

The similarity between the two mass relations
$m_dm_t\approx m^2_c \leftrightarrow m_um_b\approx m^2_s$
suggests that the mass matrices $M_{up}$ and $M_{down}$  share a
common structure. Such a desired feature
arises naturally provided $Q^{up}_{ij}\leftrightarrow Q^{down}_{ij}$
under $\alpha \leftrightarrow -\beta$ , $a \leftrightarrow -b$.
The latter, to be referred to as the
Up-Down symmetry, is violated of course (spontaneously) by
$\mid x\mid\ll\mid y\mid$ and $\mid n\mid\ll\mid m\mid$.
The various $Q$-charges get restricted by the Light-Heavy
symmetry. The latter, being manifest via $Q_{ij}=Q_{ji}$,
dictates $\alpha=b$ and $\beta=a$.
In turn, the underlying $U(1)_Q$ appears to be \underline{axial},
a feature long ago recognized \onlinecite{axial}
as a vital ingredient for flavor-chiral family grand unification.

To analyze the interplay of the $Q$-charges, let us focus attention
on $Q^{up}_{ij}$ ($Q^{down}_{ij}$ is obtained via $a \leftrightarrow -b$).
The information collected so far can be summarized by
\begin{equation}
Q^{up}_{ij}=
\left(
\begin{array}{ccc|cc|c}
.&.&.&.&.&b\\
\hline
.&.&.&.&b&.\\
.&.&.&a&p_1&q_1\\
\hline
.&.&a&.&.&.\\
.&b&p_2&.&q_2&.\\
b&p_3&.&.&.&.\\
\end{array}
\right)
{}~,
\end{equation}
where $p_1=p_2$, and $p_3$, $q_1$, $q_2$ have yet to be determined.
Once they are specified, owing to $Q_{ij}\equiv Q_i+Q_j$, all the left
over matrix elements of $Q_{ij}$ get fixed
(for example, $Q_{26}=Q_{25}-Q_{35}+Q_{36}=b-p_1+q_1$).
The locations of $q_{1,2}$
signal the block-mixing entries (see our earlier
discussion) in the corresponding $M^{up}_{ij}$, and we have allowed for
the option $p_i\neq 0$ in case the heavy mass scale $M$ also turns out to be
spontaneously generated.
To determine the $p$'s and $q$'s as linear combinations of
$a$ and $b$, we first note the crucial restriction due to
the fact that $u_{iL}$ and $d_{iL}$ form weak iso-doublets.
The latter requires $Q(u_{iL})=Q(d_{iL})$, and hence,
$Q(t_L)-Q(c_L)=Q_{16}-Q_{26}=b-(b+q_1-p_1)$
and $Q(c_L)-Q(u_L)=Q_{25}-Q_{45}=b-(a+q_2-p_2)$ must stay
invariant under $a \leftrightarrow -b$.
Few other requirements such as
$p_3=\pm p_1$  (to have $M$ entries in all $p_i$ locations),
$p_i \rightarrow \pm p_i$ under $a \leftrightarrow -b$  (to allow for the same
$M$  entries in the down sector), and $p_1+p_2-q_2 \neq a,q_1$
(to fully distinguish the right handed fermions), complete the picture.
Altogether, we derive $p_1=p_2=-p_3=-{1\over 2}(a+b)$ and $q_1=q_2=-b$.

The completion of the $Q^{up}_{ij}$  matrix allows us, by virtue of
eq.(\ref{defQ}), to finally extract the charges of the fermions themselves.
We have already arranged for $Q_{Right}(a,b)=-Q_{Left}(a,b)$ and
$Q_{down}(a,b)=Q_{up}(-b,-a)$, so we only need to specify the charges of the
left-handed up-quarks. We derive
\begin{equation}
\begin{array}{cclccclc}
Q(t_L)&=&{1 \over 2}(4b-a)&,&Q(T_L)&=&-{1\over 2}b&, \\
Q(c_L)&=&{3 \over 2}b&,&Q(C_L)&=&{1\over 2}(a-2b)&,\\
Q(u_L)&=&{3 \over 2}a&,&Q(U_L)&=&-{1\over 2}a&, \\
\end{array}
\label{Qnum}
\end{equation}
and study their implications.

Actually, corresponding to the two degrees of freedom in eq.(\ref{Qnum}), two
axial symmetries underlie our analysis:

\noindent \underline{\bf I. Flavor-blind $Z_3$}:
The first symmetry has to do with
$a=b$. This only
distinguishes a standard fermion, for which $Q(u_{iL})={3\over4}(a+b)$,
from a seesaw
fermion characterized by $Q(U_{iL})=-{1\over 4}(a+b)$. But this is not
necessarily in accord with $SU(2)_L \ast U(1)_Y$ which obviously requires
$Q(u_{iL})=Q(d_{iL})$. In other words, ${3 \over 4}(a+b)$ must
not change under $a \leftrightarrow -b$, a severe constraint
which can only be satisfied provided the symmetry in hand is $Z_3$,
such that $ e^{i{3 \over 2}(a+b)}=1$.
We note in passing that the associated anomaly vanishes.

\noindent \underline{\bf II. Horizontal $Z_5$}:
The second symmetry has to do with $a+b=0$. It is easy to
verify that the charges of the three seesaw fermions form an
arithmetic sequence, namely
$2Q(T_L)=Q(U_L)+Q(C_L)$. This may upset the Light-Heavy symmetry
principle since in general $2Q(t_L)\neq Q(u_L)+Q(c_L)$. However, in analogy
with the previous case, there exists a way out.
One can easily verify that $e^{i{5 \over 2}(a-b)}=1$ is the desired
constraint which makes $Q(u_{iL})-Q(U_{iL})$ flavor-blind, thereby
defining our horizontal $Z_5$ sub-group.

The discrete symmetry factors fit nicely in our overall
 philosophy. Reconstructing the $Z_3 \ast Z_5$  assignments of the scalars
involved, we notice that $a,b={1\over 2}(a+b)\pm{1\over 2}(a-b)$ whereas
${1\over 2}(a+b)={1\over 2}(a+b)+0\cdot{1\over 2}(a-b)$, so that $M$ (unlike
$x$,$y$) is $Z_5$-invariant. This observation can be naturally translated
into the spontaneous symmetry breaking chain
\begin{equation}
Z_3 \ast Z_5 \stackrel{M}{\longrightarrow} Z_5
\stackrel{x,y}{\longrightarrow} 1 ~,
\end{equation}
thereby constituting a group theoretical origin for the Fermi mass hierarchy.
This way, with $M$ being spontaneously rather than explicitly generated, our
model resembles, in some respects, the Majoron model \onlinecite{majoron}
in neutrino physics.
Giving up the $Z_3$-symmetry will allow $M$ to be explicitly assigned,
in analogy with the Gell-Mann-Yanagida model
\onlinecite{Gell-Mann-Yanagida}.
Furthermore, one cannot ignore the facts that the three singlet scalars share
a common $Z_3$-charge and have their $Z_5$-charges form an arithmetic
sequence. This suggests, if one is willing to introduce another scalar doublet
(with a tiny VEV), a possible extension of the family structure to
the Higgs system as well.

Adopting the so-called Yukawa universality \onlinecite{YU} as a working
hypothesis, and paying
attention to additional entries as dictated by the $Z_3 \ast Z_5$
symmetry, we obtain
\begin{equation}
M_{up}=
\left(
\begin{array}{ccc|cc|c}
0&0&0&0&0&m\\
\hline
0&0&0&0&m&0\\
0&0&x^*&x&M&y^*\\
\hline
0&0&n&0&0&0\\
0&y&M&0&y^*&0\\
y&M^*&y^*&0&0&0\\
\end{array}
\right)
{}~,
\end{equation}
accompanied by $M_{down}=M_{up}(m\leftrightarrow n^*,x\leftrightarrow y^*)$.
Note that, without any lose of generality, $M$ can be made real using
three (physically unimportant) right-handed phases.
Now, reordering the rows and columns of the above mass matrix,
we can bring it to the canonical form
$\left(\begin{array}{cc}\alpha&\beta\\ \gamma&MI+\delta\\
\end{array}\right)$, where $\alpha,\beta={\cal O}(m,n)$,
$\gamma,\delta={\cal O}(x,y)$, and the identity $I$ are 3x3 matrices.
This way it is easier to perform the perturbative expansion to
obtain the effective low-energy mass matrix $m_{eff}=\alpha-
{1\over M}\beta\gamma+{1\over M^2}(\beta\delta-
{1\over2}\alpha\gamma^{\dag})\gamma-
{1\over M^3}(\beta\delta^2-\alpha\gamma^{\dag}\delta-
{1\over2}\beta\gamma\gamma^{\dag})\gamma +... $
\begin{equation}
m^{(up)}_{eff}\approx
\left(
\begin{array}{ccc}
ny^2& nxy^*& ny^{*2}\\
0&-mx&-my^*\\
0&-{1\over 2}mxy&m(1-{1\over2}\mid y \mid^2)\\
\end{array}
\right)
{}~,
\end{equation}
where, from this point on, $x$,$y$ stand for $x \over M$,$y \over M$. The
diagonalization procedure in both quark sectors confirms the spectrum as
prescribed by eq.(\ref{spectrum}), and furthermore gives rise
to the following Cabibbo-Kobayashi-Maskawa unitary matrix
\begin{equation}
V_{CKM}\approx
\left(
\begin{array}{ccc}
1&{n\over m}y^*-{m^* \over n^*}x&x\left({m^* \over n^*}x-{n\over m}y^*\right)\\
{m\over n}x^*-{n^* \over m^*}y&1&y^*-x\\
y\left({n^* \over m^*}y-{m\over n}x^*\right)&x^*-y&1\\
\end{array}
\right)
{}~.
\end{equation}
As advertised, reflecting the hybrid nature of the Fermi hierarchy as
expressed by the $\mid{m\over n}\mid$ enhancement, $\mid V_{us}\mid \approx
{m_c \over m_b}$ (roughly 0.23 at the conventional 1GeV mass scale) is
significantly larger than the other 'nearest neighbor' $\mid V_{cb}\mid
\approx {m_s \over m_b}$ (roughly 0.04). All mixing angles, in particular the
$\mid {V_{ub} \over V_{cb}} \mid \approx {m_d\over m_s}$  ratio,
agree quite well with the experimental data, and so does the predicted
$CP$-violating invariant phase. In fact, using Wolfenstein
parameterization \onlinecite{wolf},
one finds $\tan \phi_{CP}={Im(xy) \over Re(xy)-\mid x \mid^2}$  which
can be arbitrarily large, as required, given
$\mid x\mid\ll\mid y\mid $. Such an $m,n$
- independent expression for $\phi_{CP}$ is quite intriguing, clearly
indicating
that
the origin of $CP$-violation in the present model lies really beyond the
electro/weak mass scale.

To summarize, we have attempted in this paper to account for the fermion
mass hierarchy within the framework of the Universal Seesaw mechanism,
where every standard fermion has a heavy seesaw partner.
The imposition of a Light-Heavy symmetry leads to a Diophantine equation
relating the number of families $N$ to the sum of the distinct integers
$n_i$ characterizing the hierarchy. To our surprise, this equation has
a unique non-trivial solution $N=3$, and the hierarchy is necessarily
geometric $m_W$, $m_W\epsilon$, $m_W\epsilon^2$. Using this fact,
we have constructed a model for three
quark families with a precisely defined symmetry between the up and down
sectors. Hybrid quark mass relations play then a crucial role in
deriving novel expressions for the CKM mixings \onlinecite{LNS}
in terms of simple quark mass-ratios (to be contrasted with
square-mass-ratios).
To start with, in order to provide selection rules
for the allowed entries in the mass matrix,
we have invoked an additional $U(1)$ symmetry.
However, consistency with the standard electro/weak theory
allows only for its axial $Z_3 \ast Z_5$ sub-group, supporting the
spontaneous breaking chain
$Z_3 \ast Z_5 \stackrel{M}{\longrightarrow}Z_5
\stackrel{x,y}{\longrightarrow} 1$.
Since only the ratios ${x \over m}$ and
${y\over M}$ survive in the low-energy regime, we have made no
attempt to probe the $M$-scale itself. This is why the Yukawa
universality has been invoked only at the working hypothesis level.
However, the seesaw model tells us that a typical neutrino mass is of order
${m^2 \over M}$, so it is neutrino physics which is expected
\onlinecite{DSW} to fix $M$. And finally,
recalling that the reconciliation of string unification with low-energy
may in fact require \onlinecite{Faraggi} exotic seesaw matter,
we anticipate some of the ideas
presented in this paper to find their origin in a grand unified theory.

\acknowledgments
It is our pleasure to acknowledge interesting and useful discussions with
Professors S.L. Glashow, N.K. Nielsen, Y. Achiman, E. Gedalin, and Dr. U. Paz.

\end{document}